# Plasmonic resonances in nanostructured transparent conducting oxide films

Jongbum Kim, Gururaj V. Naik, Naresh K. Emani, and Alexandra Boltasseva

*Abstract*— Transparent conducting oxides (TCO) are emerging as possible alternative constituent materials to replace noble metals such as silver and gold for low-loss plasmonic and metamaterial (MMs) applications in the near infrared (NIR) regime. The optical characteristics of TCOs have been studied to evaluate the functionalities and potential of these materials as metal substitutes in plasmonic and MM devices, even apart from their usual use as electrode materials. However, patterning TCOs at the nanoscale, which is necessary for plasmonic and MM devices, is not well-studied. This paper investigates nanopatterning processes for TCOs, especially the lift-off technique with electron-beam lithography, and the realization of plasmonic nanostructures with TCOs. By employing the developed nanopatterning process, we fabricate 2D-periodic arrays of TCO nanodisks and characterize the material's plasmonic properties to evaluate the performance of TCOs as metal substitutes. Light-induced collective oscillations of the free electrons in the TCOs (bulk plasmons) and localized surface plasmon resonances are observed in the wavelength range from 1.6µm to 2.1µm. Well-defined resonance peaks are observed, which can be dramatically tuned by varying the amount of dopant and by thermally annealing the TCO nanodisks in nitrogen gas ambient while maintaining the low-loss properties.

## I. Introduction

FOR plasmonic systems [1, 2], noble metals traditionally have been used as metallic components due to their ability to support collective oscillations of free electrons (plasmons) at optical frequencies. However, the performance of plasmonic devices has been severely limited by the large optical losses in noble metals [3, 4]. Recently, many studies have been conducted to characterize the optical properties of various alternative plasmonic materials such as metal alloys [5-7], transition-metal nitrides [8, 9], and heavily doped semiconductors [3, 10-13] in order to find materials that could outperform noble metals for plasmonic and MM applications. Such low-loss plasmonic materials would aid in overcoming the limitations of practical applications for plasmonic and MM systems. Recent studies have demonstrated that transparent conducting oxides (TCOs) such as Al- and Ga-doped zinc oxide (AZO, GZO) and tin-doped indium oxide (ITO) are good candidates as plasmonic materials in the near infrared frequency range because they exhibit metallic behavior and smaller losses compared to those of silver and gold in the NIR [8, 10, 12]. So far, there are a few demonstrations where TCOs have been used as plasmonic materials, such as in semiconductor plasmonic quantum dots [14-16], plasmonic modulators [17] and negative refraction in AZO/ZnO MMs [18]. In these demonstrations, TCO-based devices showed better performance compared to noble metal-based devices in terms of lower losses and tunability. In addition, there are many other applications such as epsilon-near-zero devices [19-21], plasmonic metasurfaces such as polarization- sensitive surfaces [22, 23] and plasmonic gas sensors [21, 24, 25] where TCOs can be better alternatives to noble metals. Here, we report on the tunable plasmonic resonances in a metasurface formed by TCO nanodisks.

TCOs have long been used in display technologies as electrode materials [26, 27] because of their transparency in the visible range and low electrical resistance. As plasmonic components, TCOs exhibit lower optical loss with small magnitudes of real permittivity compared to those parameters exhibited by noble metals. TCOs additionally offer great modulation and switching possibilities that can new generation of tunable plasmonic and MM devices. To exhibit plasmonic properties, TCOs must have a carrier concentration higher than $10^{20}$ cm$^{-3}$, which leads metal-like behavior in the NIR. Note that if the carrier concentration in the host semiconductor of the TCO does not reach this level, the material will function as a dielectric in the NIR. By careful control and optimization of the fabrication conditions such as dopant type, doping concentration, deposition temperature and deposition pressure, one can fabricate TCOs that exhibit the critical optical properties suitable for plasmonic applications in the NIR. Our recent studies showed that while being plasmonic, TCOs can have losses four times smaller than that of silver in the NIR [3].

Manuscript received August 20, 2012. This work was supported by the ONR MURI grant N00014-10-1-0942.

J. Kim is with the School of Electrical and Computer Engineering, Purdue University, West Lafayette, IN 47907-2035, USA, (e-mail: kim668@purdue.edu).
G. V. Naik is with the School of Electrical and Computer Engineering, Purdue University, West Lafayette, IN 47907-2035, USA, (e-mail: gnaik@purdue.edu).
N. K. Emani is with the School of Electrical and Computer Engineering, Purdue University, West Lafayette, IN 47907-2035, USA, (e-mail: emani@purdue.edu).
A. Boltasseva is with the School of Electrical and Computer Engineering, Purdue University, West Lafayette, IN 47907-2035, USA, (e-mail: aeb@purdue.edu).

Those two factors (low loss and plasmonic properties) open new routes to designing and realizing plasmonic and optical MM devices with high performance at the technologically important near infrared wavelengths region including the telecommunication window around 1.55 µm.

The next important step along the path to replacing conventional metals with new materials is to develop the necessary nanopatterning techniques to make the new materials into designs and devices. This is a critical step because most plasmonic and MM devices are based on building blocks of nanostructured metals and dielectrics [28-30]. In our studies, we use a lift-off process with electron-beam lithography (EBL), a commonly used method to pattern nanoscale devices, to produce 2D-periodic arrays of TCO nanodisks. The plasmonic properties of the TCO nanodisks are analyzed for a disk diameter range of 250-900 nm and for a constant disk height of 270 nm. We observe localized surface plasmon resonances (LSPRs) in the TCO nanodisk array at NIR frequencies, and we find that the LSPR wavelength and full-width-half-maxima (FWHM) of the resonance are remarkably sensitive both to the dimensions of the nanodisks and the doping density as well as to a subsequent thermal annealing treatment.

The paper is organized as follows: In Section II we describe the procedure of the lift-off process and the restrictions on the deposition conditions set by the lift-off process. The optical properties of TCOs are strongly dependent on the deposition conditions, and thus optimization is necessary to achieve the TCO properties required for plasmonic operation in the desired wavelength range. The TCO thin films are characterized with spectroscopic ellipsometry to retrieve their dielectric functions. Prism coupling experiments are performed to study the surface plasmon polariton (SPP) characteristics of the TCOs. In Section III we analyze the plasmonic properties of the arrays of TCO nanodisks and demonstrate that the LSPR absorption of GZO nanodisks can be dynamically tuned across the NIR spectrum while maintaining the low-loss properties of the TCO by applying a thermal annealing treatment.

## II. Fabrication and Characterization

### A. Lift-off Process

To fabricate a 2D array of TCO nanodisks as depicted schematically in Fig. 1, a silicon substrate was first spin-coated with a 1-µm-thick layer of positive electron-beam resist (ZEP 520A) followed by the sample pre-bake at 180 °C for 2 min. The nanoscale pattern of cylindrical nanodisks was then exposed by EBL (Vistec VB6). The beam energy was 100 kV, and the beam current was 1.012 nA. The base dose was maintained at 320 µC/cm². The exposed sample was developed in ZED-N50 (n-amyl acetate) for 1 minute, and dipped in isopropyl alcohol for 30 seconds, and then dried in gaseous nitrogen. Prior to film deposition, a post-bake was performed at 200 °C for 30 sec. We deposited TCO films by pulsed laser deposition (PVD Products, Inc.) using a KrF excimer laser (Lambda Physik GmbH) operating at a wavelength of 248 nm for source material ablation.

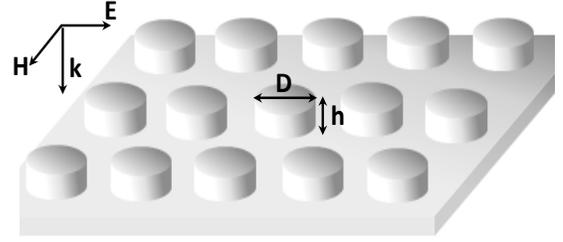

Fig. 1. Schematic view of a bi-periodic array of TCO nanodisks and the definition of the relevant parameters.

The GZO, AZO, and ITO targets were purchased from the Kurt J. Lesker Corp. with purities of 99.99% or higher. The energy density of the laser beam at the target surface was maintained at 1.5 J/cm². A high partial pressure can etch the e-beam resist during the deposition process due to reaction with oxygen [31]. Thus, all the films were grown with an oxygen partial pressure of 0.2 mTorr (0.027 Pa) or lower. Since e-beam resist can become hard-baked from elevated substrate temperatures during a deposition process, the deposition temperature should be maintained as low as possible in order to facilitate the subsequent lift-off process. In our studies, the substrate temperature during TCO thin film deposition was optimized at 70 °C.

For the lift-off process, the sample deposited with a TCO film was dipped in ZDMAC (dimethylacetamide) for 10 min and sonicated for 1 min. Most of the e-beam resist was removed during this process, but small amounts of resist remained on the edges and sides of the nanostructures. In order to remove the residual e-beam resist, the sample was dipped in PRS 2000 stripper at 70 °C for 30 min and then dipped in acetone for 5 min for rinsing.

### B. Ellipsometric Characterization

The optical properties of the TCO films were characterized by spectroscopic ellipsometry (V-VASE, J. A. Woollam) in the spectral region from 350 nm to 2000 nm. The dielectric function of the film was retrieved by fitting a Drude + Lorentz oscillator model to the ellipsometry data. The following equation (1) describes the Drude + Lorentz oscillator model.

$$\varepsilon(\omega) = \varepsilon_\infty - \frac{\omega_p^2}{\omega(\omega + i\Gamma_p)} + \sum_{m=1}^{n} \frac{f_m \omega_m^2}{\omega_m^2 - \omega^2 + i\omega\Gamma_m} \quad (n=1) \quad (1)$$

Here $\varepsilon_\infty$ is the background permittivity, $\omega_p$ is the unscreened plasma frequency, $\Gamma_p$ is the carrier relaxation rate, and $f_m$ is the strength of the Lorentz oscillator with center frequency $\omega_m$ and damping $\Gamma_m$.

Fig. 2 shows the optical properties of AZO, GZO and ITO films deposited under conditions optimized for the lift-off process. Notably, the cross-over wavelengths of all the TCO films are below telecommunication wavelength of 1.55 μm. Compared to our previous study [8], the optical properties of these TCO films are improved in terms of their metallic behavior and optical losses. The AZO film offers the lowest optical loss, referring to the imaginary part of its permittivity. Note that GZO can provide a cross-over wavelength as low as 1.2 μm, but the optical loss in GZO is higher than that in AZO. Under conditions optimized for the lift-off process, the ITO film has the highest loss and the largest cross-over wavelength compared to AZO and GZO.

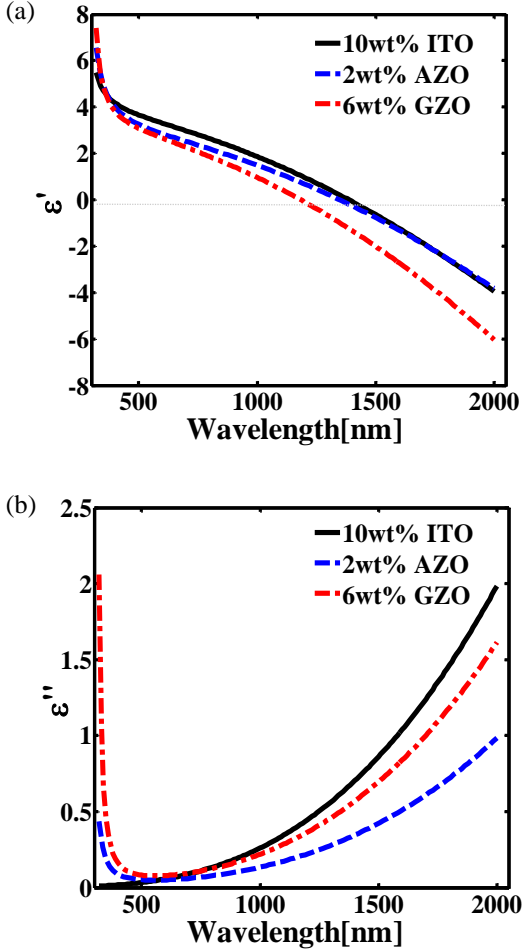

Fig. 2. Comparison of the optical properties of TCO films deposited under optimized conditions for the lift-off process. (a) Real part of the spectral dielectric function, (b) Imaginary part of spectral dielectric function.

## C. Prism Coupling for SPP Measurements

Surface plasmon polaritons are propagating charge-density waves on metal–dielectric interface that can be excited by attenuated total reflection of an incident electromagnetic wave [32]. In this work, SPP excitation on TCO films was used to verify the applicability of these materials for NIR plasmonic devices, especially at the telecommunication wavelength of 1.55 μm. We used a prism coupler (Metricon 2010/M) and implemented a Kretschmann–Raether configuration for SPP coupling (see Fig. 3). The TCO thin films were directly deposited on BK7 glass coupling prisms (n=1.501), and the thicknesses of AZO, GZO and ITO were 154 nm, 147 nm and 139 nm, respectively. A beam of TM-polarized, monochromatic laser at a wavelength of 1.55 μm was used to illuminate the sample through the input facet of the 45° BK7 glass coupling prism. While rotating the sample with respect to the laser beam, the far-field reflectance was measured with a detector. This provided a measurement of the reflected intensity for a range of internal angles from 30° to 62°. Theoretically, SPPs at a TCO-air interface are expected in wavelength region where the real part of the TCO permittivity ($\varepsilon'_{TCO}$) is less than −1.

The experimental observation of broad SPP resonances in ITO films was previously reported in [33, 34]. Those reports demonstrated a thickness-dependent SPP on ITO thin films. AZO was previously reported to be incapable of supporting SPPs at 1.55 μm because of its smaller plasma frequency. However, the AZO films in this and our previous work [8] are optimized for large plasma frequencies at 1.55 μm. The experimental data from the prism coupling reflectance measurements clearly shows the SPP existence on AZO films at a wavelength of 1.55 μm (Fig. 3(b)). The reflectance measurements from the prism coupler were verified using analytic calculations. Fig. 3(c) shows the calculated reflectance values for AZO, GZO and ITO thin films. The dip in reflectance occurring around 50-60° corresponds to the excitation of SPPs on these films.

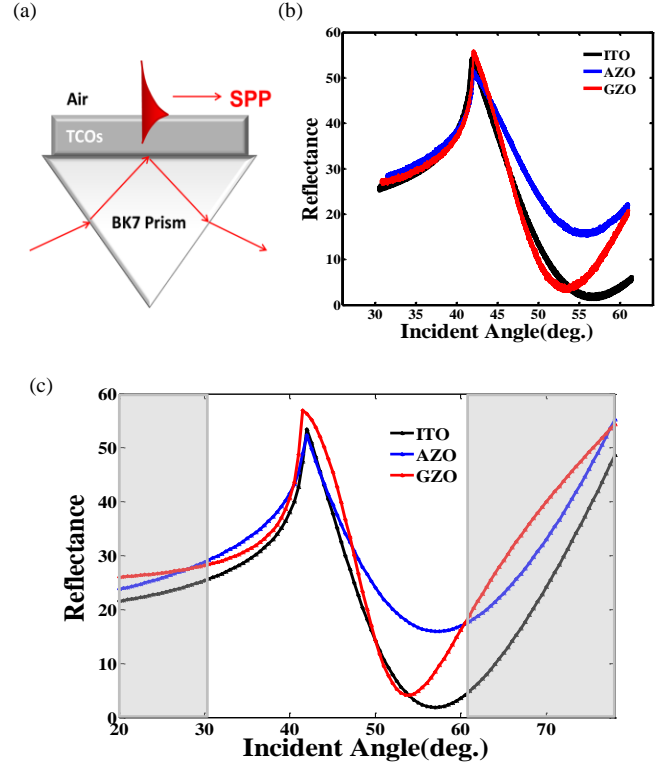

Fig. 3. (a) Schematic view of the experimental setup for SPP excitation in attenuated total reflection. (b) Reflectance curve vs. incident angle of light with 1.55 μm wavelength for ITO, AZO and GZO. (c) Simulation of reflectance curve vs. incident angle of light with 1.55μm wavelength for ITO, AZO, and GZO.

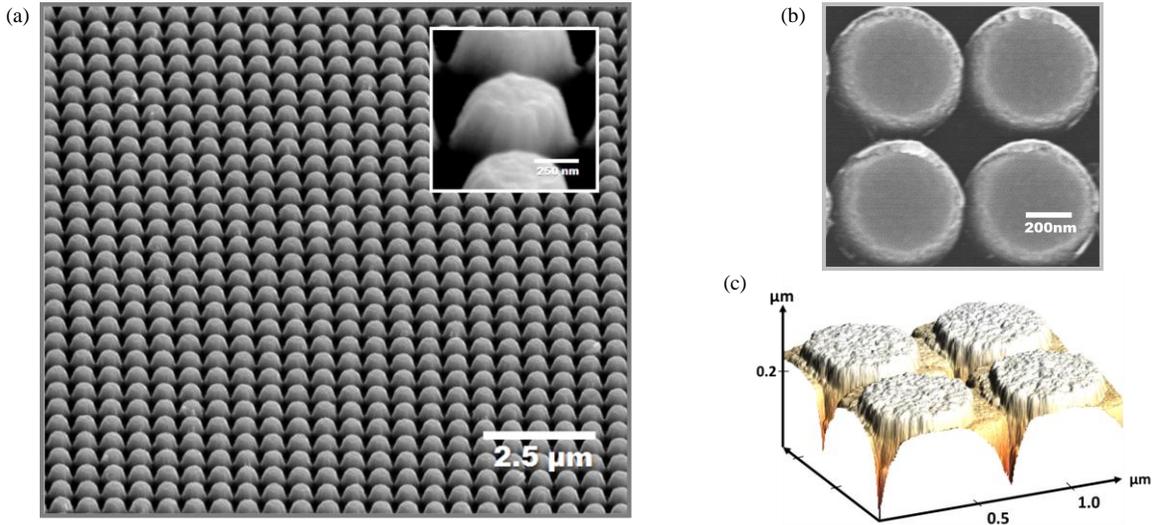

Fig. 4. (a) 54° tilted SEM image of an array of GZO nanodisks with a mean diameter D=500 nm and height h=270 nm. (inset) SEM image of GZO nanodisks at high magnification. (b) Top-down view of the nanodisks showing nearly circular shapes. (c) AFM scan of the GZO nanodisks.

## III. ARRAYS OF NANODISKS

### A. Structural Characterization

Plasmonic structures such as nanodisks of noble metals have been studied extensively [35, 36] since their strong resonant interaction with light is useful in many applications such as sensors. In this work, a polarization-independent design consisting of a periodic 2D-array of nanodisks is used to study the plasmonic properties of TCO nanostructures and to compare those properties to previous studies with noble metals. As shown in Fig. 4, we fabricated a square array of 270-nm-thick GZO nanodisks with a spacing of 100 nm between adjacent nanodisks. The nanodisk diameter was varied from 250 nm to 900 nm over a number of samples. In order to make the LSPR structures covering much of the NIR spectrum (including the telecommunications wavelengths), we fabricated the nanodisk array with GZO because it has higher plasma frequency compared to other TCOs. The scanning electron microscope (SEM) image in Fig. 4(a) shows the uniformity of the nanopatterned arrays in a relatively large area of nanoscale devices. The shape of nanodisk is almost perfectly circular shown in Fig. 4(b).

It is important to note that the deposition of the GZO layer on a patterned e-beam resist and its subsequent lift-off produces non-vertical side walls. As a result, the cross-section of the nanodisk represents a trapezoidal shape (see the inset of Fig. 4(a)). For morphological analysis, we scanned the sample with an atomic force microscope (Veeco Dimension 3100 AFM) to check the roughness of the nanodisk top surface. We used standard Si probe tips with the AFM in tapping mode. The resolutions of our AFM scans were not sufficient to accurately investigate the full depths of the narrow gaps between nanodisks. Hence, it is difficult to see the cross-sectional dimension of the nanodisks from the AFM image shown in Fig. 4(c). The root-mean-squared (RMS) roughness of the tops of the patterned nanodisks was about 6-8 nm. For as-deposited GZO thin films without any patterning processing, the RMS roughness is 5~7 nm. We can therefore confirm that the lift-off process does not significantly affect the surface morphology of the developed TCO material.

### B. Optical Characterization

The transmission spectra of the nanodisk arrays are obtained using a V-VASE spectroscopic ellipsometer with a normally incident TE wave. The measurement is performed in the wavelength range from 1.1µm to 2.4 µm (see Fig. 5). Note that absorption below 1.2 µm corresponds to phonon-assisted interband optical absorption in the silicon substrate. The LSPR wavelength and intensity depend on the size, shape, and properties of the nanostructured array [37, 38].

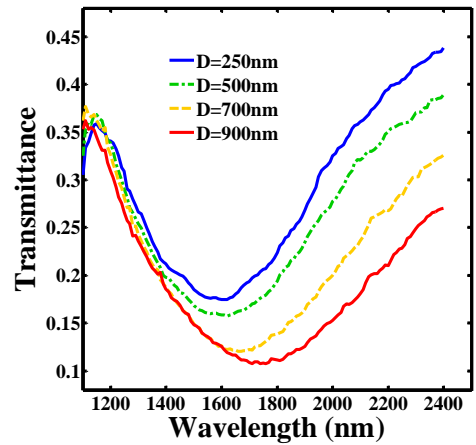

Fig. 5. Transmittance spectra for GZO nanodisk array samples on Si substrates with different nanodisk diameters.

In our studies, we investigate the effects of disk size and doping density on the LSPR properties. The transmission spectra reveal well-defined LSPR peaks, and the positions of these peaks depend on both the disk size and the doping density. As the disk diameter increases (see Fig. 5), the resonance red shifts and becomes stronger because the disks would begin to support higher order plasmonic modes. The experimental trends are verified by simulation using 3D-spatial harmonic analysis (SHA) [39]. The simulations are performed assuming ideal cylindrical nanodisks and using the optical properties obtained from thin films co-deposited with the nanostructures. Figure 6(b) shows the simulation results for the GZO nanodisks showing the changes caused by doping density. The simulations qualitatively agree with the experiments. However, there are also some deviations, which we believe are due to fabrication imperfections and the trapezoidal shape of nanodisks. In our previous study [8], we reported the change of plasma frequency and optical loss depending on doping density. As the doping density of GZO increases (see Fig. 6(a)), the films exhibit higher plasma frequency, and hence, the resonance shifts to shorter wavelengths. The optical loss of GZO is increased as reducing the doping density. The broadening of resonance peak corresponds to the increase of optical loss. In terms of the tunability of the LSPR wavelength, the peak shift arising from the change in doping density is much stronger than that caused by the nanodisk geometry.

## C. Thermal Annealing

Thermal treatments on TCO films have been well-studied in transparent electrode research in order to enhance the crystallinity and hence, the transparency of TCO films [40-42]. The effect of thermal annealing on a TCO film is strongly dependent on the temperature and the type of ambient gas. In order to characterize the effect of thermal annealing on plasmonic properties, we first investigated the annealing effect on the optical properties of TCOs with respect to two aspects: carrier concentration and optical loss. The GZO nanodisk sample was annealed up to 350 ºC for an hour in nitrogen ambient to observe the effect of the annealing gas on the optical loss.

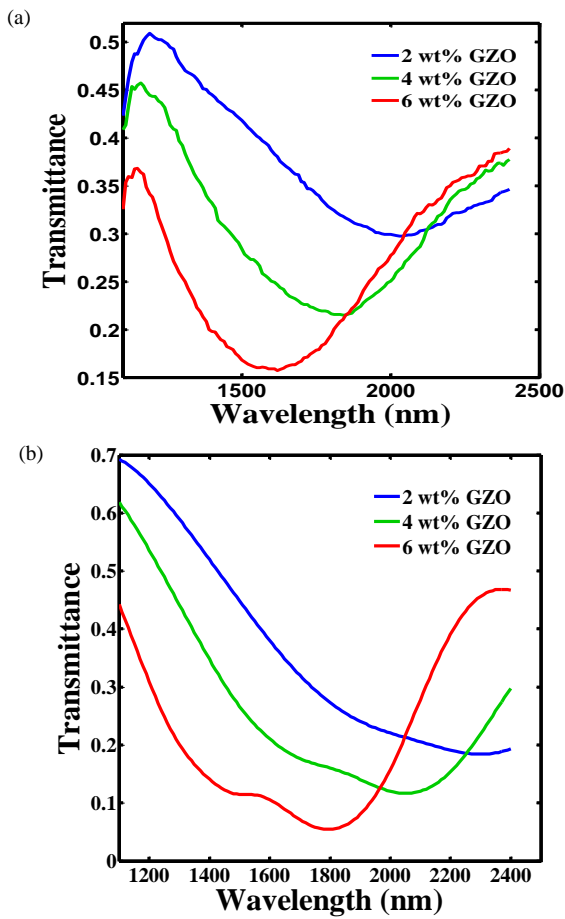

Fig. 6. (a) Measured transmittance spectra for the GZO nanodisk arrays (disk diameter of 500 nm) with different doping ratios in the GZO material. (b) Simulation results of transmittance spectra for GZO nanodisk arrays using different dielectric functions for films with different doping concentrations.

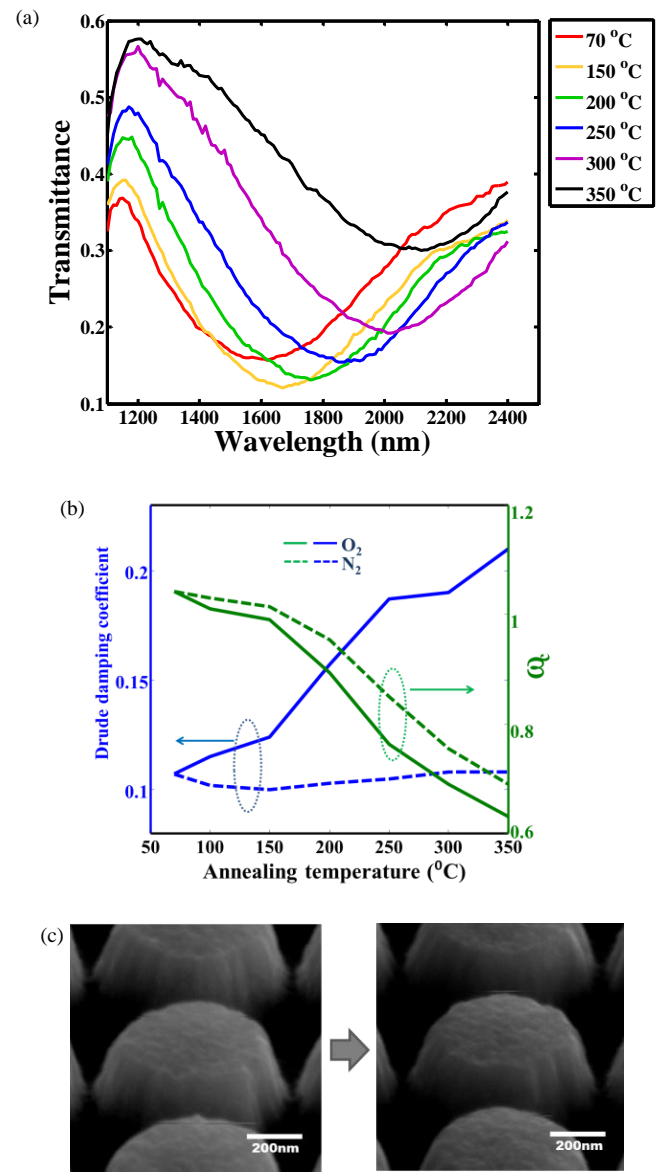

Fig. 7. (a) Transmittance spectra for GZO nanodisk arrays with different thermal annealing temperatures in a nitrogen ambient. (b) Drude damping coefficient and cross-over frequency ($\omega_c$) vs. annealing temperature in either oxygen or nitrogen ambient gas. (c) SEM image of nanodisk before and after thermal treatment.

The resulting transmittance spectra in Fig. 7(a) show that the thermal treatment can dramatically tune the LSPR peak to longer wavelengths due to reductions in the carrier concentration. Post-deposition anneal offers a way to control the LSPR properties through a post-fabrication treatment without any changes in optical loss of TCOs. This allows for flexibility in the design and optimization of the LSPR nanostructure. In Fig. 7(b) we plot the Drude damping coefficient and cross-over frequency ($\omega_c$) as functions of the annealing temperature with either a nitrogen or oxygen ambient. The Drude damping coefficient is indicative of the optical losses occurring in the material, and the cross-over frequency ($\omega_c$) is defined as the frequency at which the real part of permittivity of the material crosses zero. Since $\omega_c$ is directly proportional to the plasma frequency ($\omega_p$), and $\omega_p$ is proportional to the square of the carrier concentration, the plot in Fig. 7(b) in essence shows the carrier concentration trend with respect to the annealing temperature. We see in the figure that the carrier concentration decreases with increasing annealing temperature for both types of ambient gas. The optical loss strongly increases after annealing in oxygen ambient, while the optical loss remains the same after annealing in the nitrogen ambient.

The morphological and structural modifications incurred by the annealing treatment have already been examined in the case of noble metals [35, 43]. In those studies the goal was to improve quality of the LSPR properties through an annealing treatment. We carry out similar studies on TCOs in this work. The SEM image in Fig. 7(c) shows that there are no substantial changes in the nanodisk shape or morphology for annealing temperatures up to 350 °C. Given that TCOs are ceramics, we would expect this trend to continue for higher temperatures as well. In contrast, noble metal nanostructures are known to deform when annealed at such temperatures.

## IV. CONCLUSION

In conclusion, TCOs are good alternatives to noble metals for plasmonic applications in the NIR. We observed that thin films of AZO, GZO and ITO can support SPPs at telecommunication wavelengths. We showed that standard nanofabrication techniques may be used to pattern these TCO films. When patterned, these materials exhibit LSPR properties similar to gold and silver nanostructures. The resonance properties strongly depend on the properties of the film such as carrier concentration. Thermal annealing in different gases altered the resonance by changing the carrier concentration in these films. At the same time, in contrast to noble metals, no significant changes in morphology, surface roughness and grain structure were observed in GZO nanodisks after annealing. The effect of the carrier concentration via annealing can be used for post-fabrication tuning of the properties of TCO devices. Such tunability of the TCO properties could be used to tailor the optical resonance for various plasmonic applications and enable a new generation of controllable, switchable devices.

ACKNOWLEDGMENT

This work was supported by ONR-MURI grant N00014-10-1-0942.